\documentclass[10pt,journal,compsoc, onecolumn]{IEEEtran}
\usepackage{cite}
\usepackage[pdftex]{graphicx}
\DeclareGraphicsExtensions{.pdf,.jpeg,.png}
\usepackage{array}
\usepackage{amsmath}
\usepackage{amssymb}
\usepackage{mdwmath}
\usepackage{mdwtab}
\usepackage{eqparbox}
\usepackage{url}
\usepackage{hyperref}
\usepackage{multirow}
\usepackage{lineno} % 'switch' option for two-column documents

\begin{document}

% \linenumbers
\title{Efficient MedSAMs: Segment Anything in Medical Images on Laptop}

\author{Jun Ma, Feifei Li$^*$, Sumin Kim$^*$, Reza Asakereh, %
Bao-Hiep Le, Dang-Khoa Nguyen-Vu, Alexander Pfefferle, Muxin Wei, Ruochen Gao, Donghang Lyu, Songxiao Yang, Lennart Purucker, Zdravko Marinov, Marius Staring, Haisheng Lu, Thuy Thanh Dao, Xincheng Ye, Zhi Li, %
Gianluca Brugnara, Philipp Vollmuth, Martha Foltyn-Dumitru, Jaeyoung Cho, Mustafa Ahmed Mahmutoglu, Martin Bendszus, Irada Pflüger, Aditya Rastogi, Dong Ni, Xin Yang, Guang-Quan Zhou, Kaini Wang, Nicholas Heller, Nikolaos Papanikolopoulos, Christopher Weight, Yubing Tong, Jayaram K Udupa, Cahill J. Patrick, Yaqi Wang, Yifan Zhang, Francisco Contijoch, Elliot McVeigh, Xin Ye, Shucheng He, Robert Haase, Thomas Pinetz, Alexander Radbruch, Inga Krause, Erich Kobler, Jian He, Yucheng Tang, Haichun Yang, Yuankai Huo, Gongning Luo, Kaisar Kushibar, Jandos Amankulov, Dias Toleshbayev, Amangeldi Mukhamejan, Jan Egger, Antonio Pepe, Christina Gsaxner, Gijs Luijten, Shohei Fujita, Tomohiro Kikuchi, Benedikt Wiestler, Jan S. Kirschke, Ezequiel de la Rosa, Federico Bolelli, Luca Lumetti, Costantino Grana, Kunpeng Xie, Guomin Wu, Behrus Puladi, Carlos Martín-Isla, Karim Lekadir, Victor M. Campello, Wei Shao, Wayne Brisbane, Hongxu Jiang, Hao Wei, Wu Yuan, Shuangle Li, %
Yuyin Zhou, and Bo Wang % <-this % stops a space

}

\IEEEtitleabstractindextext{%
\begin{abstract}
Promptable segmentation foundation models have emerged as a transformative approach to addressing the diverse needs in medical images, but most existing models require expensive computing, posing a big barrier to their adoption in clinical practice. In this work, we organized the first international competition dedicated to promptable medical image segmentation, featuring a large-scale dataset spanning nine common imaging modalities from over 20 different institutions. The top teams developed lightweight segmentation foundation models and implemented an efficient inference pipeline that substantially reduced computational requirements while maintaining state-of-the-art segmentation accuracy. Moreover, the post-challenge phase advanced the algorithms through the design of performance booster and reproducibility tasks, resulting in improved algorithms and validated reproducibility of the winning solution. Furthermore, the best-performing algorithms have been incorporated into the open-source software with a user-friendly interface to facilitate clinical adoption. The data and code are publicly available to foster the further development of medical image segmentation foundation models and pave the way for impactful real-world applications.
\end{abstract}

}

\maketitle

\IEEEdisplaynontitleabstractindextext

% up to 2000 words 

% For peer review papers, this IEEEtran command inserts a page break and
\IEEEpeerreviewmaketitle

\section*{Introduction}
% segmentation and the recent foundation models
Segmentation is a fundamental task in medical image analysis, aiming to provide accurate boundaries for anatomies and pathologies, which plays an important role in many clinical tasks, such as disease detection and diagnosis, surgical planning, and treatment monitoring~\cite{heartNature23,pancreasAI-NMed}. Over the past decade, deep learning-based models have revolutionized medical image segmentation by providing state-of-the-art accuracy and unprecedented automation~\cite{seg-reviewPAMI}. More recently, the emergence of foundation models, such as Segment Anything Model (SAM)~\cite{2023-SAM1-Meta,SAM2}, has introduced a new paradigm for segmentation tasks~\cite{SAM1-Eval-Maciej,SAM1-Eval-XinYang}.
These models are trained on large-scale annotated images and capable of being generalized across multiple domains and transferred to unseen domains~\cite{2023-SEEM}. For example, MedSAM~\cite{MedSAM} and BiomedParse~\cite{BiomedParse} have demonstrated strong zero-shot segmentation ability across common medical imaging modalities and a wide range of anatomies by fine-tuning SAM on over one million medical image-mask pairs.

% bottleneck: hard to deploy on laptop; lack of benchmark
However, despite these advances, adopting foundation models for segmentation in clinical practice remains an ongoing challenge. One key bottleneck preventing the widespread use of these foundation models is the high demand for computing resources. 
Many of these segmentation foundation models are resource-intensive because of large model weights, making them difficult to deploy on commonly available hardware, such as laptops, creating a barrier to democratizing the use of advanced segmentation tools, especially in resource-constrained clinical environments. Additionally, as segmentation foundation models have been developed and validated under different settings, it is difficult to assess their performance rigorously. This calls for standardized benchmarks, yet their absence hinders the identification of optimal models~\cite{InterSegSurveyTPAMI}.

% international challenge can help stimulate method development
International competitions have proven to be effective in driving advancements in various domains, such as ImageNet competition~\cite{Imagenet} for deep convolutional neural networks~\cite{AlexNet} and the critical assessment of protein structure prediction competition for AlphaFold~\cite{AlphaFold1}. By providing a standard platform for researchers to compare their models, these challenges foster collaborative efforts of researchers all over the world to identify novel approaches while ensuring transparent evaluation and reproducible results~\cite{NatMed-panda,FLARE22,touchstone}. Thus, an international benchmark for promptable medical image segmentation can greatly boost the progress in creating lightweight, efficient models that are also suitable for clinical deployment.

% contribution of this work
In this work, we present the competition results for promptable medical image segmentation, specifically designed to encourage the development of more efficient segmentation foundation models for diverse modalities and segmentation targets.
Importantly, we collaborate with 24 institutions worldwide to curate a new testing set with over 4000 cases, allowing for fair model evaluation while minimizing the risk of data leakage. Based on this comprehensive benchmark dataset, the top-performing algorithms demonstrate substantial improvements in both segmentation accuracy and efficiency, with models achieving segmentation results over ten times faster than existing SAM-based foundation models~\cite{2023-SAM1-Meta,MedSAM}. Moreover, we work with the two best-performing algorithm developers and integrated the models into the open-source image computing platform 3D Slicer with a user-friendly interface, making state-of-the-art segmentation models more accessible and practical for a wider range of clinical applications.

\begin{figure}[htbp]
\centering
\includegraphics[scale=0.26]{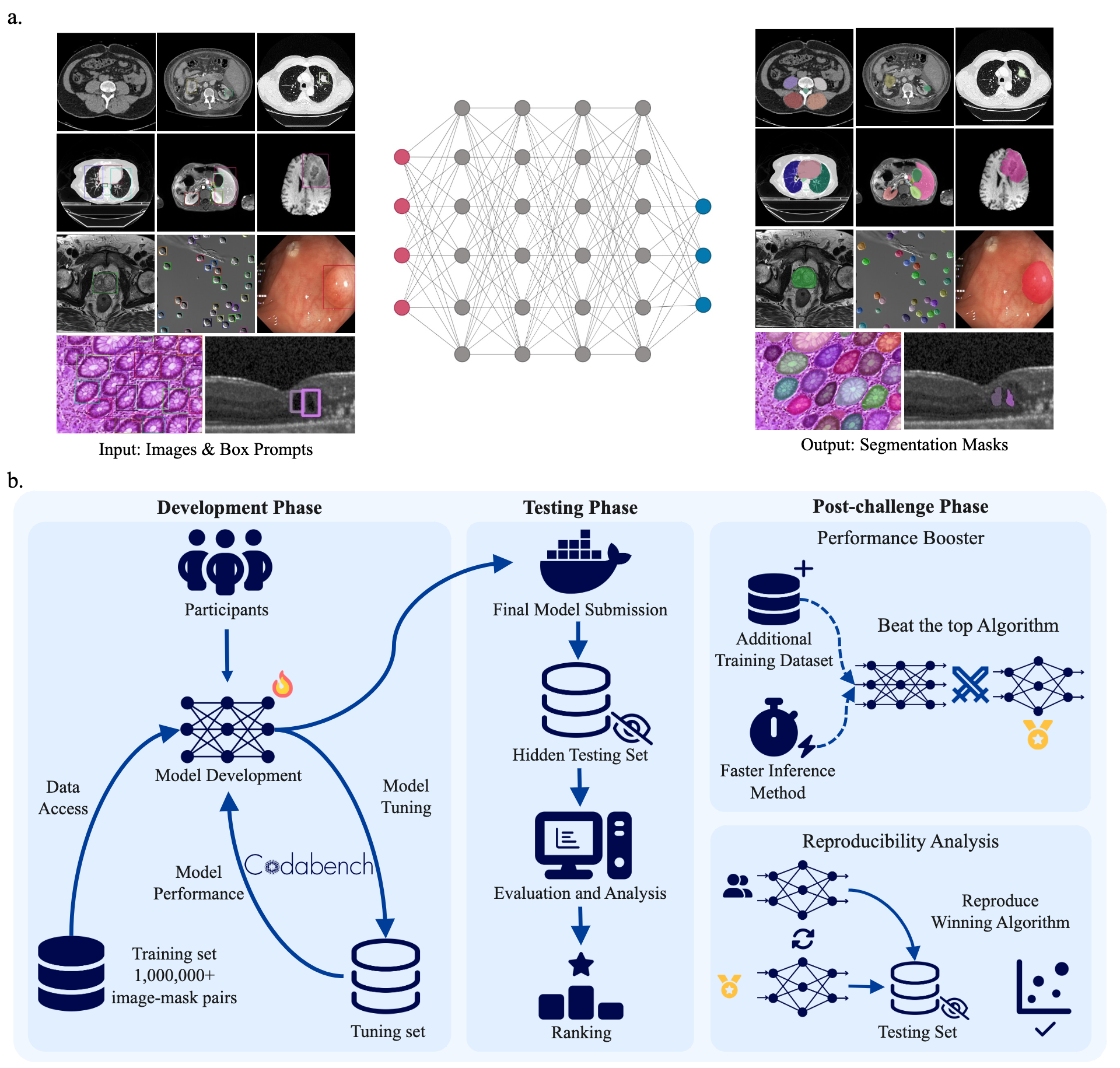}
\caption{\textbf{Competition design.} \textbf{a,} The task is to develop universal segmentation foundation models that can accept various medical image inputs with target bounding box prompts and generate the corresponding segmentation masks. The model should be lightweight and deployable on laptops without reliance on graphics processing unit (GPU). \textbf{b,} Three phases in the competition. During the development phase, participants train their models on the training set and obtain performance metrics on the online validation set on Codabench. The top 20 teams on the validation leaderboard are invited to submit their algorithm dockers and we manually evaluate them on the hidden testing set.
After that, we release all the technical details and code of the top ten teams and launch a post-challenge phase to invite participants to further boost their model performance and reproduce the winning algorithms.
}\label{fig:1}
\end{figure}

\section*{Results}
% data set introduction
\subsection*{Competition Design: Universal and Lightweight Medical Image Segmentation Foundation Models} 
The competition task is designed for promptable medical image segmentation. Specifically, the input contains various medical images and bounding boxes for segmentation targets, and the output is the corresponding segmentation mask. The bounding box is selected as the prompt because it can precisely specify the target, which has lower ambiguity than point prompts.  
% We assign a unique bounding box for each target with a random perturbation on the coordinates (Methods). 
The challenge consists of three distinct phases: a development phase lasting 122 days, a testing phase spanning 35 days, and a post-challenge phase of 35 days. These phases focus respectively on model training, evaluation on a hidden testing set, and subsequent improvements and reproducibility analysis.

During the development phase, participants were provided with a large-scale and diverse training dataset, serving as the foundation for participants to design, develop, and train their models.
Moreover, an online leaderboard was made available on Codabench~\cite{codabench}, allowing for automatic evaluation of participants’ results. Participants could leverage this feature to test their models against a tuning set, obtain performance feedback, and iteratively tune and refine their approaches. This phase was instrumental in enabling participants to optimize their models and prepare them for subsequent testing.

During the testing phase, the top 20 teams from the validation leaderboard were invited to make testing submissions. However, submissions were not limited to these top teams and other participants were also welcome to submit their solutions for evaluation. Participants were required to encapsulate their algorithms in Docker containers, ensuring a standardized and portable format for evaluation. Each Docker container was manually executed on a hidden testing set, with all evaluations conducted on the same workstation to ensure fairness and consistency. The teams were ranked based on both the accuracy and efficiency of their models. Specifically, the evaluation criteria included the Dice Similarity Coefficient (DSC), Normalized Surface Distance (NSD), and runtime performance. This rigorous and equitable process determined the testing leaderboard rankings.

The post-challenge was designed with two primary objectives: further improving model performance by integrating successful strategies from the top-performing algorithms (performance booster) and evaluating the reproducibility of the winning solutions (reproducibility analysis). This phase aimed to push the boundaries of segmentation performance using collective knowledge and assess whether the winning solution could be easily implemented by others. In the performance-booster subtask, participants were encouraged to enhance their models by incorporating two new datasets, fast inference methods, and other optimization strategies contributed by participants. 
In the reproducibility subtask, all participants were invited to reproduce the results of the top-performing teams. We made the code and corresponding methodologies of top solutions publicly available, allowing participants to validate them on the testing set. These collaborative efforts not only fostered the development of a refined solution capable of outperforming the original winning algorithms but also verified the reproducibility of the top solutions.

% leveraging a newly aggregated training dataset, which combined the original training set with two additional external datasets provided by participants.

\begin{figure}[htbp]
\centering
\includegraphics[scale=0.43]{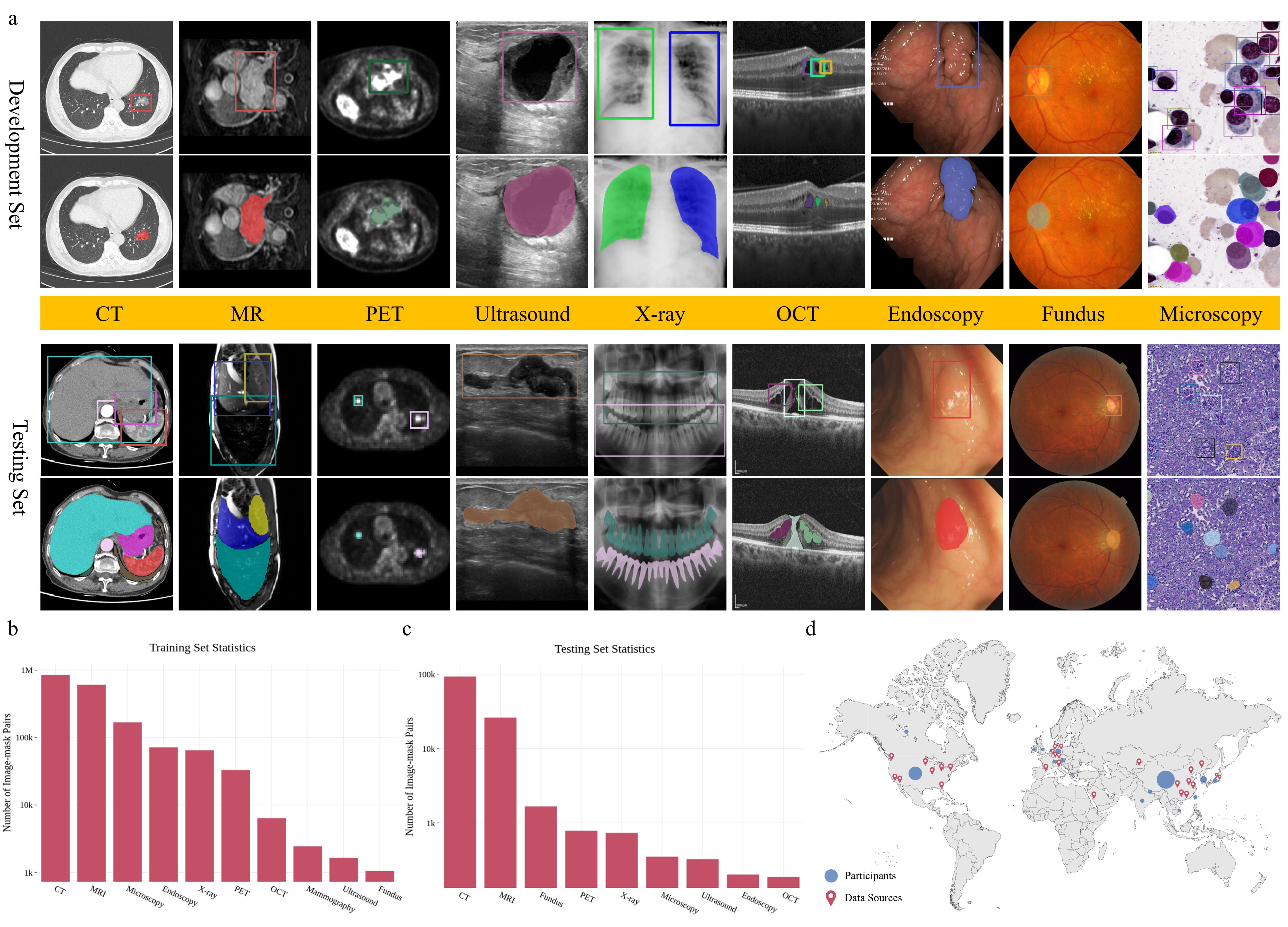}
\caption{\textbf{Competition dataset.} \textbf{a,} Example images and the annotations in the training and testing set. 
\textbf{b,} The number of image-mask pairs in the public training set. 
\textbf{c,} The number of image-mask pairs in the hidden testing set. All testing data has been newly collected specifically for this challenge and was not publicly available beforehand.
\textbf{d,} Geographical distribution of participants and data contributors in the testing set. 
}\label{fig:2}
\end{figure}

\subsection*{Competition Dataset: A Global Collaboration}
We provided participants with a diverse set of training and testing images to facilitate robust model development and evaluation. Figure~\ref{fig:2}a illustrates several visual examples from the training and testing sets across various imaging modalities, accompanied by their corresponding bounding box prompts and reference masks. 
The training dataset comprises 1,809,644 image-mask pairs spanning 10 imaging modalities, including Computed Tomography (CT), Magnetic Resonance Imaging (MRI), Positron Emission Tomography (PET), ultrasound, X-ray, Optical Coherence Tomography (OCT), endoscopy, fundus, and microscopy images. Figure~\ref{fig:2}b visualizes the distribution of image-mask pairs across these modalities, with CT and MRI images dominating the dataset due to their widespread use in clinical diagnostics.
All images were sourced from publicly available datasets with proper licensing permissions and were preprocessed to ensure standardization in intensity range and format. However, some public datasets prohibit redistribution. To address this limitation, we compiled a comprehensive list of over 100 medical image segmentation datasets, detailing information such as modality, segmentation targets, number of cases, and download links. Participants were also encouraged to contribute by adding new public datasets to the list.
To ensure a fair comparison, participants were restricted to using only the officially preprocessed data and datasets included in the curated list for model development.

Since the majority of the public medical image segmentation datasets have been included in this competition, we opted to create a completely new testing set rather than annotating existing datasets to mitigate the potential risk of data leakage.   To ensure the dataset was both diverse and representative, we launched a “Call for Dataset” initiative, inviting researchers from around the globe to contribute de-identified medical images in accordance with predefined criteria and licensing requirements (Methods).
As a result, we successfully collected 124,004 image-mask pairs from 4,414 cases (Fig.~\ref{fig:2}c, Supplementary Table 3). While the testing set spans a wide range of imaging modalities, CT and MRI images continue to dominate, reflecting their prevalence in clinical diagnostics. 

Fig.~\ref{fig:2}d illustrates the geographical distribution of participants and testing data sources in the competition. The blue circles represent the locations of participant teams, with their size corresponding to the relative number of teams in each region. The challenge attracted participation from over 200 individuals and 61 data contributors across 24 institutions, which highlights the international interest in advancing medical image segmentation and demonstrates the competition's success in engaging a wide array of contributors from different regions around the world.

\subsection*{Algorithm Overview}
We first developed a baseline model, LiteMedSAM, by distilling the large vision transformer encoder~\cite{ViT2020ICLR} in MedSAM to a TinyViT~\cite{TinyViT} (Methods). The code and model were publicly available at the beginning of the competition to reduce the entrance barrier for participants.  
In addition to the baseline model, we received 22 algorithm submissions during the testing phase where 20 algorithms were from the top 20 teams on the validation leaderboard. 

Most of the top-performing teams employed a SAM-like promptable segmentation model, including an image encoder to extract features from input images, a prompt encoder to process the bounding box coordinates, and a mask decoder to generate segmentation masks. The key to standing out is to design lightweight image encoders and implement fast inference strategies. 
We have summarized the key components of the top five algorithms, focusing on their network architectures and efficient inference strategies, in Supplementary Table 4. In this section, we specifically highlight the top three best-performing algorithms.

\textbf{Best-performing algorithm.} Le et al. (T1-seno~\cite{CVPR24-EffMedSAMs-1st}) proposed MedficientSAM, which used the EfficientViT-L1 model as the image encoder and conducted knowledge distillation from the MedSAM encoder~\cite{MedSAM} on the training set. Then, the whole model was fine-tuned in an end-to-end way where the prompt encoder and the mask decoder were initialized with MedSAM's pretrained weights. For faster inference, the Python-based pipeline was ported to C++, and the trained model was exported to OpenVINO format. Moreover, an embedding caching mechanism was implemented for 3D images, which only needed to compute image embeddings once for multiple box prompts. Furthermore, all the modules were compiled from source code to take advantage of runtime optimizations such as Advanced Vector Extensions and Link Time Optimization.

\textbf{Second-best-performing algorithm.} Pfefferle et al. (T2-automlfreiburg~\cite{CVPR24-EffMedSAMs-2nd}) introduced a data-aware fine-tuning framework to efficiently produce tailored models for specific data or modality types. The image encoder was the lightweight EfficientViT-L0 model with a three-stage fine-tuning pipeline: knowledge distillation, general fine-tuning on the whole training set, and data-aware fine-tuning on a subset. In addition to the above OpenVINO and embedding caching mechanism for inference optimization, model caching and slim packages (e.g., Python 3.11 and OpenCV headless version) were used to speed up loading models and dockers, respectively.

\textbf{Third-best-performing algorithm.} Wei et al. (T3-skippinglegday~\cite{CVPR24-EffMedSAMs-3rd}) proposed RepMedSAM, which replaced MedSAM's image encoder with a pure CNN and lightweight RepVit-M model~\cite{RepViT}. The prompt encoder and mask decoder were from MedSAM, while the image encoder was trained by knowledge distillation. No specific runtime optimization was implemented during inference because the designed model already had relatively better latency compared to the baseline model. 

\begin{figure}[htbp]
\centering
\includegraphics[scale=0.4]{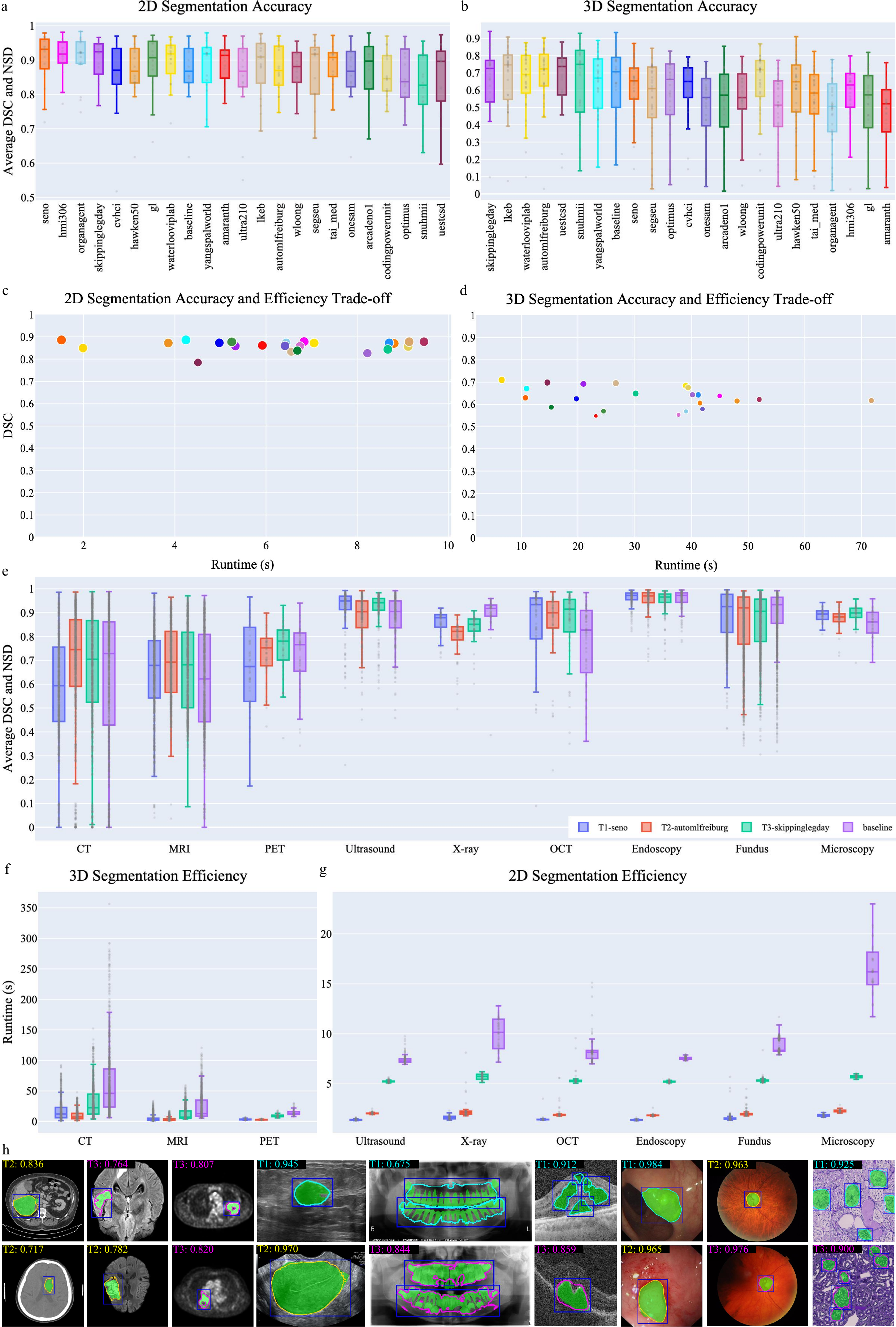}
\end{figure}

\begin{figure}[htbp]
\caption{\textbf{Evaluation results of 23 algorithms on the hidden testing set.} 
Dot and box plot of the Average DSC and NSD scores on the \textbf{a,} 2D (n = 2,309 images) and \textbf{b,} 3D (n = 2,105 scans) testing set. The box plots display descriptive statistics across all testing cases, with the median value represented by the horizontal line within the box, the lower and upper quartiles delineating the borders of the box and the vertical black lines indicating 1.5 × IQR. The algorithms are organized on the x-axis based on their corresponding ranks.
The bubble plots show the trade-off between segmentation accuracy (DSC and NSD) and efficiency on the \textbf{c,} 2D and \textbf{d,} 3D testing set. The circle size is proportional to the NSD score.
\textbf{e,} Modality-wise segmentation accuracy performance (Average DSC and NSD) of three best-performing teams and LiteMedSAM baseline. 
Modality-wise segmentation efficiency (runtime) of three best-performing teams and LiteMedSAM baseline on 3D \textbf{f,} and 2D \textbf{g,} modalities.
\textbf{h,} Visualized segmentation results of nine modalities. The blue bounding box and green overlay denote prompt and reference standard, respectively. For each image, the best DSC score from the three algorithms and the corresponding contour are presented.
%note: the text in f should be normalized.
}\label{fig:3}
\end{figure}

\subsection*{Performance analysis on the hidden testing set}
All the submitted algorithms were independently executed on the same computing platform (Methods) for a fair comparison. 
Fig~\ref{fig:3}a and b show the average DSC and NSD scores across the 2D and 3D datasets, respectively. The 2D segmentation results reveal generally high average DSC and NSD scores for most algorithms, with several teams, such as seno and hmi306, achieving median scores over 0.9. Those teams also have relatively narrow interquartile ranges (IQRs), suggesting that their algorithms demonstrate both high accuracy and consistent performance across different cases.

The overall range of scores for 3D segmentation appears broader than in the 2D case, reflecting the added complexity of volumetric segmentation tasks.
3D segmentation results also exhibit more variability with substantially lower median scores compared to 2D results. While teams such as skippinglegday and waterlooviplab maintain strong performance with high average DSC and NSD scores, many teams display wider IQRs and longer whiskers, indicating greater inconsistency in their segmentation results. 

In addition to segmentation accuracy, the efficiency of the segmentation algorithm is another crucial aspect, as it substantially impacts the overall user experience. The bubble plots in Fig~\ref{fig:3}c and d present the trade-off between segmentation accuracy and efficiency, measured as runtime in seconds, for the 2D and 3D datasets, respectively. The size of the circles is proportional to the NSD score. For 2D segmentation, most algorithms achieve high DSC scores (above 0.8) while maintaining a runtime of less than 10 seconds per image. Notably, teams seno and automlfreiburg achieve exceptional accuracy and efficiency, with a NSD near 0.9 and a runtime less than 2 seconds. In contrast, 3D segmentation results show a wider spread in runtimes, ranging from approximately 10 seconds to over 70 seconds per case, reflecting the greater computational complexity of volumetric data processing. Nevertheless, teams seno and automlfreiburg still obtain the fastest inference with competitive DSC and NSD scores, demonstrating a great trade-off between segmentation accuracy and efficiency.  

% modality-wise analysis
Next, we conducted a fine-grained analysis of the three overall best-performing teams and baseline across the nine modalities (Fig~\ref{fig:3}e, Supplementary Table 5-9). For the 3D modalities CT, MRI, and PET, T2-automlfreiburg and T3-skippinglegday achieve comparable performances, which are significantly better than T1 ($p<0.01$). Across most 2D modalities, T1-seno consistently outperforms the others, achieving the highest accuracy with relatively narrow interquartile ranges, reflecting its robustness and consistency. T2-automlfreiburg, while competitive on endoscopy, fundus, and microscopy, shows inferior performance on the other modalities, particularly in ultrasound and X-ray. 
T3-skippinglegday approaches the performance of T1-seno for most modalities, especially excelling in microscopy. 
The LiteMedSAM baseline, while competitive for some modalities, such as PET and X-ray, lags substantially behind for most modalities, with lower scores and wider performance ranges.

Furthermore, we compare the runtime of the four algorithms across 3D and 2D modalities. For 3D modalities (Fig~\ref{fig:3}f), T1-seno and T2-automlfreiburg demonstrate consistently lower runtimes with minimal variance than other algorithms, indicating their efficiency and stability across all 3D images. In contrast, the baseline exhibits higher runtimes and substantial variability, particularly for CT, with runtimes exceeding 300 seconds in large CT scans. 
Similar trends are observed across 2D modalities (Fig~\ref{fig:3}g). Notably, T1-seno consistently consumes around one to two seconds for all 2D images, which is five to ten times faster than the baseline model. 

Finally, we visualized some segmentation examples of the three best-performing algorithms for each modality (Fig~\ref{fig:3}h, Supplementary Fig. 1). The top algorithms demonstrate superior versatility, achieving accurate results for most of the images even if the object boundaries are not clear. However, certain challenging cases highlight areas where further refinement is needed. For instance, in 3D modalities like CT and MR, the segmentation of heterogeneous lesions, characterized by irregular shapes and varying intensities, leads to some over- or under-segmentations. Additionally, unseen or less common targets, such as teeth segmentation in X-ray images, pose unique challenges where the algorithms struggle to generalize effectively, resulting in suboptimal predictions. These challenges could be addressed by enriching the training dataset through community-driven efforts to include more diverse and rare cases. Moreover, integrating interactive refinement mechanisms, which allow users to provide real-time corrections, could also help improve segmentation results.

%%%%%%%%%%%%%%%%%%%%%%%%%%%%%%%%%%%%%%%%%%%%%%%%%%%%%%%%%%%%%%%%%%%%%%%%%
%%%%%%%%%%%%%%%%%%%%%%%%%%%%%%%%%%%%%%%%%%%%%%%%%%%%%%%%%%%%%%%%%%%%%%%%%
%%%%%%%%%%%%%%%%%%%%%%%%%%%%%%%%%%%%%%%%%%%%%%%%%%%%%%%%%%%%%%%%%%%%%%%%%

\begin{figure}[htbp]
\centering
\includegraphics[scale=0.43]{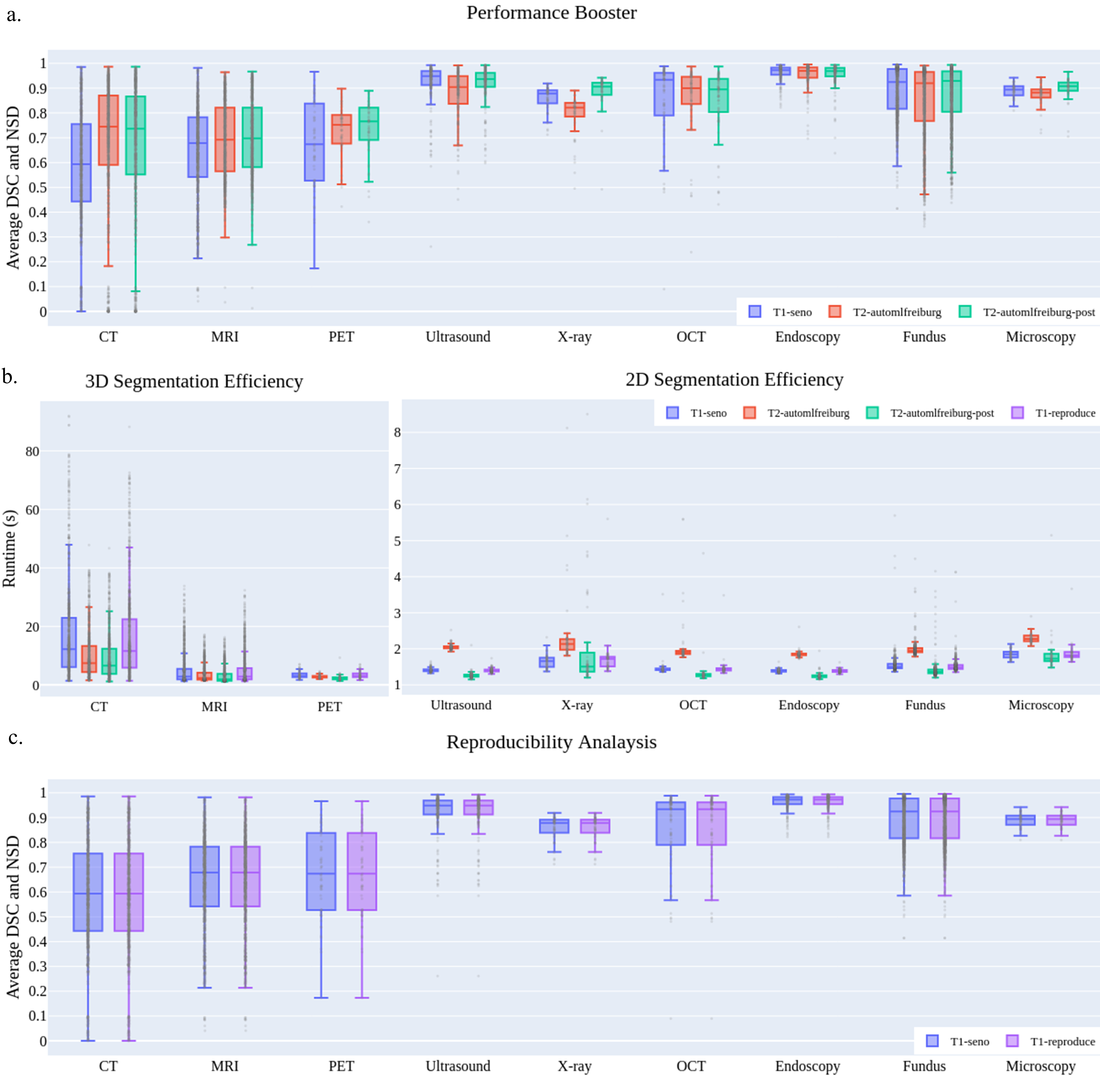}
\caption{\textbf{Results of the post-challenge.} 
\textbf{a,} The new best-performing algorithm (T2-automlfreiburg-post) and comparison to its predecessor (T2-automlfreiburg) and the previous best-performing algorithm (T1-seno) across all modalities.
\textbf{b,} Runtime comparison. 
\textbf{c,} T1-reproduce successfully replicated the performance of T1-seno across all modalities
}\label{fig:4}
\end{figure}

\subsection*{Post-challenge analysis}
We found that none of the algorithms consistently achieved the best performance across all modalities, indicating that these algorithms could be complementary. This motivated us to design a post-challenge to invite participants to enhance their algorithms by using the strategies from top solutions. Although all the top-three algorithms only used the provided training set, using external public datasets from the predefined dataset list was allowed and encouraged. To mitigate the effects of training datasets, we reached out to the teams with the highest DSC and NSD scores in each modality and identified the employed external datasets. Only one X-Ray~\cite{SciData-dentalXray} and one PET dataset~\cite{autoPET-NMI} were used as additional datasets by the modality-wise winning teams. Thus, we also added those two datasets to the training set.

We received six submissions for the performance-booster task. 
One algorithm (T2-automlfreiburg-post)~\cite{CVPR24-EffMedSAMs-2nd} surpassed the previous best-performing algorithm (T1-seno) and its previous algorithm (T2-automlfreiburg) in terms of the overall rank. The main technical improvements include using a shared EfficientViT model for all 3D modalities, early stopping to avoid over-fitting, and speeding up the inference pipeline with C$++$ implementations from T1-seno~\cite{CVPR24-EffMedSAMs-1st}.
Fig~\ref{fig:4}a shows the modality-wise performance of the three algorithms. T2-automlfreiburg-post substantially improves the previous algorithm on PET, ultrasound, X-ray, and microscopy modalities while preserving the performance for the other modalities. Importantly, the runtime is reduced by 2x, matching the efficiency of the fastest algorithm T1-seno (Fig~\ref{fig:4}b). 

In addition, we received one submission (T1-reproduce) for the reproducibility analysis task, aiming to achieve similar performance to T1-seno using the released code~\cite{CVPR24-EffMedSAMs-1st-reproduce}. The results, as shown in Fig~\ref{fig:4}c, indicate that T1-reproduce successfully replicated the performance of T1-seno across all modalities, achieving nearly identical average DSC and NSD scores, although the training batch size was smaller than the original setting because of the computing resource limitation.

\section*{Discussion}
% challenge achievement
This work presents the first international competition dedicated to promptable medical image segmentation foundation models. One of the key achievements of the competition was the curation of a large-scale, unpublished dataset for benchmarking, consisting of diverse imaging modalities and newly annotated cases. This dataset not only mitigated the risk of data leakage from existing public sources but also provided a robust and comprehensive foundation for evaluating segmentation algorithms. The challenge attracted substantial global participation, with thousands of submissions on the online leaderboard from over 200 participants, showcasing the high level of interest in the community. 

% key designs
This competition also introduced a novel task focused on efficient segmentation, aiming to develop deployable segmentation models that balance high accuracy with practical runtime performance without reliance on expensive GPUs. 
Although all the top teams employed SAM-like architectures, two key designs made their algorithms stand out: efficient network architecture and fast inference strategies. In particular, all the three best-performing algorithms replaced the heavy-weight ViT image encoder~\cite{ViT2020ICLR} in MedSAM with a lightweight image encoder, such as EfficientViT~\cite{EfficientViT2023CVPR} and RepViT~\cite{RepViT}, which reduced the parameters by 50\%. Moreover, knowledge distillation was an effective way to transfer knowledge from larger, more accurate models to compact models.
Notably, using lightweight image encoders with knowledge distillation preserved segmentation accuracy, which achieved comparable to or even better accuracy than MedSAM while substantially reducing computational overhead (Supplementary Fig. 2). 

In addition to lightweight image encoders, we also identified several practical strategies for fast inference on CPU, including embedding caching, the use of a C++ inference pipeline, and integration with OpenVINO combined with optimized Docker deployment. Embedding caching minimizes redundant computations by storing pre-computed 2D slice embeddings for 3D image segmentation. The C++ inference pipeline improves execution speed by leveraging a compiled language’s low-level optimizations, enabling faster runtimes compared to Python. Additionally, OpenVINO and optimized Docker deployment streamline inference by utilizing a highly efficient runtime environment and lightweight containerization, reducing resource consumption and startup latency. Together, these strategies enable faster and more resource-efficient deployments on edge devices without sacrificing segmentation accuracy.

Furthermore, our post-challenge phase also introduced two innovative tasks: the performance booster and the reproducibility analysis, aimed at advancing algorithm performance and ensuring scientific rigor. 
One of the new submissions set a new state-of-the-art by combining elements of existing successful solutions, demonstrating the power of collaborative advancements. The reproducibility analysis further validated the robustness of the competition’s outcomes, as the winning solution during the testing phase was successfully reproduced using the released code and documentation. This highlights the transparency and reliability of the methods developed during the challenge, ensuring that the results can serve as a solid foundation for future research.

While the competition showcased substantial algorithm advancements for universal medical image segmentation models, there remains a substantial gap between these cutting-edge algorithms and their integration into clinical practice. The complexity of deploying advanced models, combined with the need for user-friendly interfaces, often hinders their adoption in real-world settings. To bridge this gap, we collaborated with the two best-performing teams to incorporate their algorithms into 3D Slicer~\cite{Slicer}, a widely-used open-source platform for medical image analysis. This integration allows clinicians and researchers to access state-of-the-art segmentation algorithms without requiring any coding expertise, making these powerful tools more accessible and practical for routine use.

This work also has several limitations. First, while we provided one of the largest and most diverse datasets to date, the majority of curated testing images originated from North America, Europe, and Asia, with a noticeable lack of representation from South America, Africa, and Oceania. Second, all the top-performing algorithms utilized 2D models for 3D data segmentation, which overlooked the slice-wise or 3D contextual information inherent in volumetric medical images~\cite{SAM2}. This limitation may impact segmentation performance, particularly for tasks requiring spatial consistency across slices. Finally, the prompts were based on the bounding box because of its effectiveness and less ambiguity. However, this setting did not consider more flexible interactive segmentation approaches, allowing for iterative user feedback refinement. 

Future iterations of the competition will aim to address current limitations and push the boundaries further. In particular, we will expand the dataset by incorporating more images from recent public datasets~\cite{AbdomenAtlas} and under-represented regions to enhance the global generalizability and fairness of the algorithms~\cite{faireness}. Additionally, we will design a new task to benchmark interactive segmentation~\cite{InterSegSurveyTPAMI} and text-based segmentation methods, reflecting the evolving research directions~\cite{SAT-Yao,BiomedParse}. This benchmark will assess how effectively algorithms can integrate user feedback through interactive processes and respond to natural language descriptions for segmentation tasks. 

In conclusion, this competition marked a pivotal step forward in advancing promptable medical image segmentation, introducing a novel task that emphasized efficiency alongside accuracy and a large and diverse testing set from 24 different institutions.  The competition attracted over 200 participants worldwide, and the two best-performing algorithms, featuring a transformer-based lightweight image encoder and fast inference implementations, achieved over 10 times faster inference speed than the previous foundation model while maintaining comparable or even superior accuracy.
The innovative post-challenge phase further validated the reproducibility of top-performing solutions and set a new state-of-the-art through integrated strategies, highlighting the importance of transparency and collaboration in driving progress. The two best-performing algorithms have been incorporated into 3D Slicer to bridge the gap between algorithmic innovation and clinical deployment. We believe these efforts will pave the way for more robust, accessible, and impactful medical image segmentation technologies, ultimately benefiting global healthcare practices.

\subsection*{Affiliations}
\begin{itemize}
    \item Jun Ma is with AI Collaborative Centre, University Health Network; Department of Laboratory Medicine and Pathobiology, University of Toronto; Vector Institute, Toronto, Canada
    \item Feifei Li is with Peter Munk Cardiac Centre, University Health Network, Toronto, Canada. ($^*$Equal contribution)
    \item Sumin Kim is Toronto General Hospital Research Institute, University Health Network; Department of Computer Science, University of Toronto; University Health Network; Vector Institute, Toronto, Canada. ($^*$Equal contribution)
    \item Reza Asakereh is with Peter Munk Cardiac Centre, University Health Network, Toronto, Canada. 
    \item Bao-Hiep Le is with University of Science, Vietnam National University, Ho Chi Minh City, Vietnam.
    \item Dang-Khoa Nguyen-Vu is with University of Science, Vietnam National University, Ho Chi Minh City, Vietnam.
    \item Alexander Pfefferle is with Institute of Computer Science, University of Freiburg, Freiburg, Germany.

    \item Muxin Wei is with the School of Medicine and Health, Harbin Institute of Technology, Harbin, China.
    \item Ruochen Gao is with the Division of Image Processing, Department of Radiology, Leiden University Medical Center, Leiden, the Netherlands.
    \item Donghang Lyu is with the Division of Image Processing, Department of Radiology, Leiden University Medical Center, Leiden, the Netherlands.
    \item Songxiao Yang is with the Department of System and Control Engineering, School of Engineering, Institute of Science Tokyo (formerly Tokyo Institute of Technology), Tokyo, Japan.
    \item Lennart Purucker is with the Institute of Computer Science, University of Freiburg, Freiburg, Germany.
    \item Zdravko Marinov is with Institute for Anthropomatics and Robotics, Karlsruhe Institute of Technology, Karlsruhe, Germany.
    \item Marius Staring is with the Division of Image Processing, Department of Radiology, Leiden University Medical Center, Leiden, the Netherlands.
    \item Haisheng Lu is with the School of Information and Communication Engineering, University of Electronic Science and Technology of China, Chengdu, China.
    \item Thuy Thanh Dao is with the School of Electrical Engineering and Computer Science, University of Queensland, Brisbane, Australia.
    \item Xincheng Ye is with the School of Electrical Engineering and Computer Science, University of Queensland, Brisbane, Australia.
    \item Zhi Li is with the School of Cyberspace, Hangzhou Dianzi University, Hangzhou, China.
    
    \item Gianluca Brugnara is with the Division for Computational Radiology and Clinical AI and the Department of Neuroradiology, University Hospital Bonn, Germany.
    \item Philipp Vollmuth is with the Division for Computational Radiology and Clinical AI and the Department of Neuroradiology, University Hospital Bonn, Germany.
    \item Martha Foltyn-Dumitru is with the Division for Computational Radiology and Clinical AI and the Department of Neuroradiology, University Hospital Bonn, Germany.
    \item Jaeyoung Cho is with the Division for Computational Radiology and Clinical AI, the Department of Neuroradiology, University Hospital Bonn, Germany.
    \item Mustafa Ahmed Mahmutoglu is with the Department of Neuroradiology, Heidelberg University Hospital, Heidelberg, Germany.
    \item Martin Bendszus is with the Department of Neuroradiology, Heidelberg University Hospital, Heidelberg, Germany.
    \item Irada Pflüger  is with the Department of Neuroradiology, Heidelberg University Hospital, Heidelberg, Germany.
    \item Aditya Rastogi is with the Division for Computational Radiology and Clinical AI, Department of Neuroradiology, University Hospital Bonn, Germany.
    \item Dong Ni is with the School of Biomedical Engineering, Shenzhen University, Shenzhen, China.
    \item Xin Yang is with the School of Biomedical Engineering, Shenzhen University, Shenzhen, China.
    \item Guang-Quan Zhou is with the School of Biological Science and Medical Engineering, Southeast University, Nanjing,  China.
    \item Kaini Wang is with the School of Biological Science and Medical Engineering, Southeast University, Nanjing,  China.
    \item Nicholas Heller is with the Department of Urology, Cleveland Clinic, Cleveland, United States.
    \item Nikolaos Papanikolopoulos is with the Department of Computer Science, University of Minnesota, Minneapolis, United States.
    \item Christopher Weight is with the Department of Urology, Cleveland Clinic, Cleveland, United States.
    \item Yubing Tong is with the Department of Radiology, University of Pennsylvania, Philadelphia, United States.
    \item Jayaram K Udupa is with the Department of Radiology, University of Pennsylvania, Philadelphia, United States.
    \item Cahill J. Patrick is with the Department of Orthopaedic Surgery, Children's Hospital of Philadelphia, Philadelphia, United States.
    \item Yaqi Wang is with the College of Media Engineering, Communication University of Zhejiang, Hangzhou, China.
    \item Yifan Zhang is with Hangzhou Dental Hospital Group, Hangzhou, China.
    \item Francisco Contijoch is with Bioengineering, Radiology, UC San Diego, La Jolla, California, United States.
    \item Elliot McVeigh is with Bioengineering, Radiology, and Cardiology, UC San Diego, La Jolla, California, United States.
    \item Xin Ye is with the Department of Ophthalmology, Hangzhou, China.
    \item Shucheng He is with the Department of Ophthalmology, Hangzhou, China.
    \item Robert Haase is with the Department of Neuroradiology, University Hospital Bonn, Bonn, Germany.
    \item Thomas Pinetz is with the Institute for Applied Mathematics, University of Bonn, Bonn, Germany.
    \item Alexander Radbruch is with the Department of Neuroradiology, University Hospital Bonn, Bonn, Germany.
    \item Inga Krause is with the Department of Neuroradiology, University Hospital Bonn, Bonn, Germany.
    \item Erich Kobler is with the Department of Neuroradiology, University Hospital Bonn, Bonn, Germany.
    \item Jian He is with the Department of Nuclear Medicine, Nanjing Drum Tower Hospital, Nanjing, China.
    \item Yucheng Tang is with the Healthcare and Life Science, NVIDIA, Redmond, United States.
    \item Haichun Yang is with the Department of Pathology, Microbiology and Immunology, Vanderbilt University Medical Center, Nashville, United States.
    \item Yuankai Huo is with the Department of Computer Science, Vanderbilt University, Nashville, United States.
    \item Gongning Luo is with the Department of Computer Science, King Abdullah University of Science and Technology, Thuwal, Kingdom of Saudi Arabia.
    \item Kaisar Kushibar is with the Faculty of Mathematics and Computer Science, University of Barcelona, Barcelona, Spain.
    \item Jandos Amankulov is with the Radiology and Nuclear Medicine, Kazakh Institute of Oncology and Radiology, Almaty, Kazakhstan.
    \item Dias Toleshbayev is with the Radiology and Nuclear Medicine, Kazakh Institute of Oncology and Radiology, Almaty, Kazakhstan.
    \item Amangeldi Mukhamejan is with the Radiology and Nuclear Medicine, Kazakh Institute of Oncology and Radiology, Almaty, Kazakhstan.
    \item Jan Egger is with the Institute for AI in Medicine, University Hospital Essen, Essen, Germany.
    \item Antonio Pepe is with the Institute of Computer Graphics and Vision, Graz University of Technology, Graz, Austria.
    \item Christina Gsaxner is with the Institute of Computer Graphics and Vision, Graz University of Technology, Graz, Austria.
    \item Gijs Luijten is with the Institute for AI in Medicine, University Hospital Essen, Essen, Germany.
    \item Shohei Fujita is with the Department of Radiology, the University of Tokyo, Tokyo, Japan.
    \item Tomohiro Kikuchi is with the Data Science Center, Jichi Medical University, Tochigi, Japan.
    \item Benedikt Wiestler is with the AI for Image-Guided Diagnosis and Therapy, Technical University of Munich, Munich, Germany.
    \item Jan S. Kirschke is with the Department of Diagnostic and Interventional Neuroradiology, Technical University of Munich, Munich, Germany.
    \item Ezequiel de la Rosa is with the Department of Quantitative Biomedicine, University of Zurich, Zurich, Switzerland.
    \item Federico Bolelli is with the Department of Engineering, University of Modena and Reggio Emilia, Modena, Italy.
    \item Luca Lumetti is with the Department of Engineering, University of Modena and Reggio Emilia, Modena, Italy.
    \item Costantino Grana is with the Department of Engineering, University of Modena and Reggio Emilia, Modena, Italy.
    \item Kunpeng Xie is with the Department of Oral and Maxillofacial Surgery and Institute of Medical Informatics, University Hospital RWTH Aachen, Aachen, Germany.
    \item Guomin Wu is with the Department of Oral, Plastic, and  Aesthetic Surgery, School and Hospital of Stomatology, Jilin University, Changchun, China.
    \item Behrus Puladi is with the Department of Oral and Maxillofacial Surgery and Institute of Medical Informatics, University Hospital RWTH Aachen, Aachen, Germany.
    \item Carlos Martín-Isla is with the Department of Mathematics and Computer Science, University of Barcelona, Barcelona, Spain.
    \item Karim Lekadir is with the Catalan Institution for Research and Advanced Studies, University of Barcelona, Barcelona, Spain.
    \item Victor M. Campello is with the Department of Mathematics and Computer Science, University of Barcelona, Barcelona, Spain.
    \item Wei Shao is with the Department of Medicine, University of Florida, Gainesville, United States.
    \item Wayne Brisbane is with the Department of Urology, University of California, Los Angeles, Los Angeles, United States.
    \item Hongxu Jiang is with the Department of Electrical and Computer Engineering, University of Florida, Gainesville, United States.
    \item Hao Wei is with the Department of Biomedical Engineering, the Chinese University of Hong Kong, Hong Kong SAR, China.
    \item Wu Yuan is with the Department of Biomedical Engineering, the Chinese University of Hong Kong, Hong Kong SAR, China.
    \item Shuangle Li is with the Department of Ophthalmology, Zigong First People’s Hospital, Zigong, China.

    \item Yuyin Zhou is with the Department of Computer Science and Engineering, University of California, Santa Cruz, USA
    \item Bo Wang (Corresponding Author) is with Peter Munk Cardiac Centre, University Health Network; Department of Laboratory Medicine and Pathobiology and Department of Computer Science, University of Toronto; Vector Institute, Toronto, Canada. E-mail: bowang@vectorinstitute.ai 
\end{itemize}

\section*{Methods}

\subsection*{Dataset curation and pre-processing}
All the training images were from publicly available datasets with a license for re-distribution (Supplementary Table 1-2). The original images have a wide range of format, such as nifti, dicom, mhd, nrrd, jpg, and png. We normalized them to the same npz format with image and reference standard inside it, allowing participants to get rid of tedious data cleaning and focus on model development. 

The pre-processing followed common practice~\cite{nnunet21,MedSAM}. Specifically, for CT images, we first adjusted the intensity to the proper window level and width followed by rescaling to $[0, 255]$. For MRI and PET images, we applied intensity cut-off with the lower-bound and upper-bound of 0.5\% and 99.5\% percentile of foreground intensity and then rescaled the intensity to $[0, 255]$. For the 2D modalities, we applied the same intensity preprocessing as MRI if their intensity ranges are not in $[0, 255]$. Otherwise, no preprocessing was applied. For the reference standard, we converted the lesion semantic mask to instance mask, allowing the generation of lesion-wise bounding box prompts.

\subsection*{Baseline: LiteMedSAM}
We provided an out-of-the-box baseline model, LiteMedSAM, with a detailed tutorial to reduce the entry barriers. 
The development of LiteMedSAM involves a two-stage process: distillation and fine-tuning. In the first stage, we distill the extensive medical imaging knowledge from MedSAM's heavyweight ViT-B\cite{ViT2020ICLR} image encoder into Tiny-ViT\cite{TinyViT}, a compact hybrid architecture combining Transformer and convolution layers. In the second stage, we fine-tune the entire LiteMedSAM model.

To prepare the input images, for MedSAM's ViT-b encoder, the images are first resized to 1024 x 1024 using bi-cubic interpolation, while for Tiny-ViT, the longest edge of the input images is first resized to 256 using bilinear interpolation, and then the images are padded to 256 x 256 with zeros on the right and bottom to maintain a consistent size. After resizing, min-max normalization is applied to scale the intensity values from the range of [0, 255] to [0, 1]. Data augmentation is then performed, which includes random horizontal and vertical flips with a probability of 0.5. These preprocessing and augmentation steps are consistent across both the distillation and fine-tuning stages.

During the distillation process, MedSAM's image encoder serves as the teacher model, while Tiny-ViT acts as the student model. The teacher model remains frozen throughout the distillation. The output feature embeddings of both image encoders have a dimensionality of 64 x 64 x 256. We adopt Mean Squared Error (MSE) loss to encourage the feature embeddings of the student to match those of the teacher. The distillation is performed using the AdamW optimizer\cite{adamW} with a learning rate of 5e-5, a weight decay of 0.01, and a batch size of 8. The process continues until either 1000 epochs are reached or the loss no longer decreases.

In the second stage, we plug in the distilled Tiny-ViT as the image encoder and fine-tune the entire LiteMedSAM model, including the prompt encoder and mask decoder. During the fine-tuning process, a loss function comprised by the unweighted sum of the Dice loss and the binary cross-entropy (BCE) loss is adopted. The same AdamW optimizer settings as in the distillation stage are used. For fine-tuning, the input images are resized to 256 x 256 using the same image processing steps as in the distillation stage for Tiny-ViT. The fine-tuning process stops after 1000 epochs or when the training loss falls below 0.005, whichever occurs first. Both the distillation and fine-tuning of LiteMedSAM are performed on the same dataset used to train MedSAM.

\subsection*{Deployment: 3D Slicer plugin}
Despite the open sourcing of these advanced algorithms, a significant challenge remains in integrating them seamlessly into clinical workflows, as this often demands basic coding skills. To address this, we incorporated the two best-performing algorithms as a plugin for 3D Slicer, an open-source software platform for medical image analysis and three-dimensional visualization. The plugin was implemented using the ``loadable script module'' type, allowing for the easy integration of Python scripts and seamless interaction with the Slicer framework.

The module provides a user-friendly graphical interface where users can specify preprocessing options and define the region of interest (ROI) for segmentation. Its architecture is based on a generic abstract class to facilitate the easy integration of new segmentation models. Importantly, all these models can be executed on laptops without reliance on GPUs.

\subsection*{Evaluation metrics and platform}
The evaluation metrics contained two segmentation accuracy metrics and one efficiency metric. 
We followed the recommendations in Metrics Reloaded~\cite{metric-reload} to evaluate the segmentation accuracy. Specifically, we used Dice Similarity Coefficient (DSC) and Normalized Surface Distance (NSD) to quantitatively evaluate the region overlap and boundary similarity, respectively.  The efficiency was measured by runtime (seconds) and we recorded the Docker container execution time for each case. 
The submitted Docker algorithms were evaluated on the same platform with Ubuntu 20.04 system, which contains one Intel CPU (Xeon W-2133 3.60GHz) and an upperbound of 8GB RAM.
Docker version 20.10.13
The evaluation code and platform information were released at the beginning of the competition for a transparent evaluation.

\subsection*{Ranking scheme and statistical analysis}
We employed the commonly used rank-then-aggregate scheme to obtain the final rank~\cite{bakas2018brats,FLARE22}, which contains three steps. First, we computed the three metrics (DSC, NSD, and runtime) for each testing case and each algorithm. Then, we ranked the 23 algorithms teams for each modality and each metric. Finally, we averaged all the rankings to obtain the final rank. 
Wilcoxon signed-rank test was used to statistically compare the performance of the algorithms. We also used the bootstrapping (N=1000) to analyze the ranking stability.  
The analysis was based on the widely used toolkit ChallengeR~\cite{ChallengeR}.

\subsection*{Data availability}
All the datasets in this study have been publicly available on the codabench~\cite{codabench} competition website (Dataset page) \url{https://www.codabench.org/competitions/1847/}. 

\subsection*{Code availability}
The code and trained models are publicly available at \url{https://github.com/bowang-lab/MedSAM/tree/LiteMedSAM}. 3D slicer plugin can be accessed at
\url{https://github.com/bowang-lab/MedSAMSlicer}. The code and model weights of best-performing teams are available at the competition website (Ranking page) \url{https://www.codabench.org/competitions/1847/}.

\subsection*{Acknowledgements} 
This work was supported by the Natural Sciences and Engineering Research Council of Canada (RGPIN-2020-06189 and DGECR-2020-00294) and CIFAR AI Chair programs. This research was enabled, in part, by computing resources provided by the Digital Research Alliance of Canada. 
B.L. and D.N. were supported by Vingroup Innovation Foundation (VINIF) in project code VINIF.2019.DA19. 
S.Y. was supported by JST SPRING, Japan Grant Number JPMJSP2106.
L.P. was supported by the Deutsche Forschungsgemeinschaft (DFG, German Research Foundation) – Project-ID 499552394 – SFB 1597.
L.Z. was supported by Zhejiang Provincial Natural Science Foundation of China under Grant No. LDT23F01015F01.
M.A.M. was supported by Else Kr\"oner Research College for young physicians (ref. number: 2023\_EKFK.02)
N.H. was supported by DoD HT9425-23-1-0918.
Y.T., J.K.U., and C.J.P. were supported by NIH R01HL150147.
F.C. was supported by NIH K01HL143113.
E.M. was supported by NIH R01HL144678.
E.K. was supported by DFG KO 7162/1-1 543939932 and FWF 10.55776/COE12.
H.Y. was supported by DOD HT9425-23-1-0003.
K.K. was supported by Juan de la Cierva fellowship by the Ministry of Science and Innovation of Spain with reference number FJC2021-047659-I.
F.B., L.L., and C.G. were supported by the University of Modena and Reggio Emilia and Fondazione di Modena, through the FAR 2024 and FARD-2024 funds (Fondo di Ateneo per la Ricerca).
We also appreciate all the data owners for providing public medical images to the community.

\bibliographystyle{IEEEtran}
\bibliography{main-ref}

\end{document}